# Safety technology for train based on multi-sensors and braking system

Dongsoo Har

**Abstract**

This work deals with integration and transmission of safety information for smart railway vehicles and control and design of cooperative emergency brake systems for high speed trains. Due to increased speed of high-speed train, safety of passengers is becoming more critical. From this perspective, three different approaches to ensure the safety of passengers are useful. These approaches are based on integrated use of multi-sensors and emergency brake system. Methodology for integrated use of sensors to obtain situation-aware safety related information and for enhanced braking is to be discussed in details in this work

## I. Introduction

Securing safety of passengers is a very important task to train operating company. To date, many different approaches have appeared. In this work, two approaches are introduced. First approach is based on integrated use of multiple sensors. These sensors provide location information with much more increased precision [1]-[4]. The second approach is making use of air spoiler. These approaches can be used with other methods to provide elevated level of safety to passengers.

Recent trend of railway train development can be characterized in several aspects : high speed, infotainment, intelligence in driving, and so on. In particular, trend of high speed in driving is prominent and competition for high speed amongst several techno-savvy countries is becoming severe. To achieve high speed, engines or motors are distributed over multiple vehicles of train to provide increased motive power, while a single engine or motor has been mostly used for conventional trains. Increased speed and more complicated powertrain system naturally incur much higher chance of massive accidents. From this perspective, importance of proactive safety control before accident takes place cannot be over-emphasized. To implement proactive safety control requires situation-aware integration and transmission of safety information obtained from IoT sensors [5]. Types of critical IoT sensors depend on situational conditions. Thus, integration and transmission of safety information should be performed with IoT sensors providing the safety information proper for faced situation. This work is to devise a methodology how to operate IoT sensor network enabling proactive safety control for railway vehicles.

The braking performance of trains suffers mainly due to the limited friction coefficient between the steel wheel and the steel rail, and the braking distance rapidly grows with the vehicle speed (e.g. 2816m at 300km/h). Such a long braking distance poses a serious problem in an emergency even when hazardous situations can be foreseen via safety information network. In order to mitigate this problem, an emergency brake system such as eddy current brake (ECB) has been widely used. However, the braking performance of ECB system reduces at higher vehicle speed, and the ECB systems are mostly controlled in an on/off manner for emergency conditions, thus cannot be precisely controlled for normal braking conditions.

In this work, air spoilers coordinated with existing brake systems for improved high-speed braking performance and vehicle stability are proposed. The air spoilers are widely used in aircrafts for increased braking performance and downforce, but its braking force rapidly degrades at lower speed. Since existing brake systems such as ECB, regenerative brake, and friction brakes show different performance and operational characteristics, the spoiler should be carefully coordinated with the existing systems depending on braking conditions.

## II. Safety Technology Based on Multi-sensors and Air Spoiler
### A. Safety Technology Based on Multi-sensors

Approach to be attempted can be described twofold : situation-aware IoT sensor(+actuator) data processing and data transmission. When a train begins its operation, it needs to perform self-diagnosis to ensure initial safety control. When it is in operation (in driving), various types of IoT sensors for monitoring collect data to proactively perform safety control. See Fig.1. Position of train is obtained by reader on bottom of vehicle(s) receiving position information from tag on track, or by other alternative means. If the immediate segment of rail from current position is curved, gyroscopic or other type of tilt sensor monitors the tilting condition more often and proper tilting actuation is executed when possible. In addition, as the speed of the train is rising, rate of data collected by video sensor for pantograph operation and wheel/axle defect sensors is to be increased. On the other hand, interior IoT sensors, such as humidity sensor and fire sensor, for regular monitoring can take constant or close to constant data collection rates.

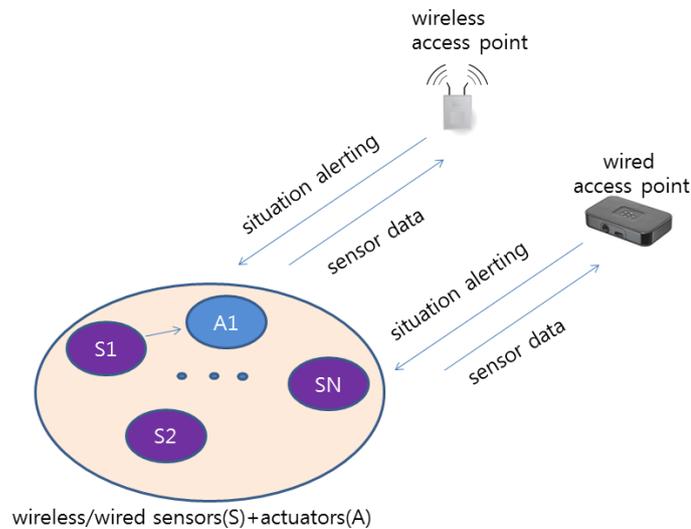

**Figure 1.** Collection rate of IoT sensor data is dynamically changed for each IoT sensor, depending on type of faced situation

Communication network within a vehicle can be formed in a combination of wired and wireless connections of IoT sensors. Some of the IoT sensor data are utilized for actuators without data transmission to external network and other data collected from sensors are transmitted to external network. To deliver the IoT sensor data in an orderly manner, a single or multiple gateways of a train to outside communication networks, e.g., LTE-R cellular network, can be placed. Access to the gateway from each sensor can be direct or hierarchical. In a hierarchical structure, cluster head which is a forwarding node to gateway can be considered as an intermediate node representing a vehicle. The sensor data are transmitted to the cluster head in each vehicle and forwarded to the gateway(sink) for eventual data aggregation. This architecture is beneficial when the number of sensors is large and coverage of whole sensor network is also large to some extent. Channel access for IoT sensor network can be prioritized according to data type and the priority of sensor data depends on situation even when they are obtained from the same sensor. Safety-critical IoT sensor data can be immediately transmitted to the external network by assigning a higher priority in contention or special time slots or even a separate gateway [6]. See Fig.2. This type of situation-aware data transmission scheme is also important. Various configurations and channel access schemes are tested to find out optimal solution.

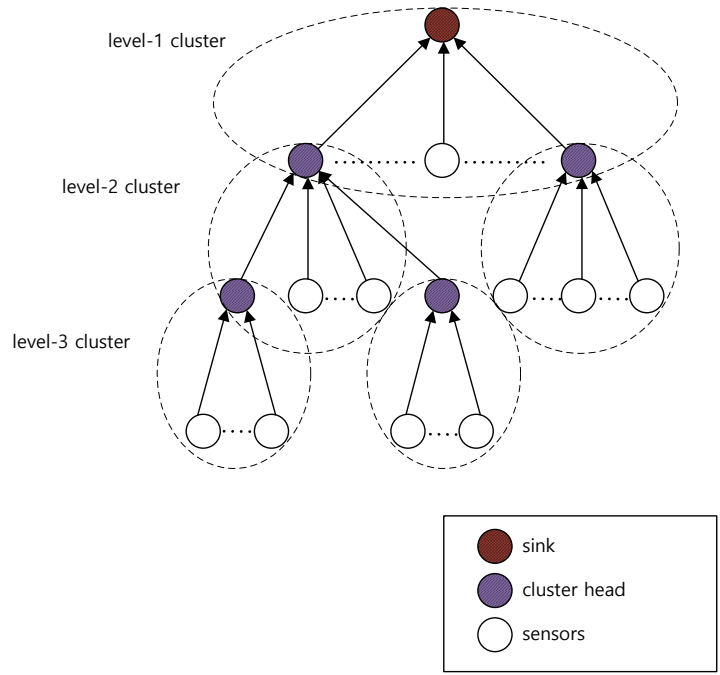

**Figure 2.** Hierarchical IoT sensor network. IoT sensors can take time-varying priority in channel access to meet required low latency.

**B. Safety Technology Based on Air Spoiler**

The goal of this braking system is to ensure the safety of high speed trains by improving braking performance and stability. To achieve this goal, design and control problems are introduced as described below.

First part of this work is the design optimization problem. There are several types of spoiler (e.g. macro geometric spoiler, micro geometric spoiler, counter-flow fluidic spoiler). Identifying characteristics of spoiler brake type and finding the best type of spoiler for train applications must be investigated [7]. The design of spoiler should also be optimized to maximize air resistance while noise and vibration must be kept minimal with sufficient controllability. Second part of this work is the optimal control and blending strategy problem. The brake systems such as ECB, regenerative brake, and friction brakes show different performance and operational characteristics. For instance, the regenerative brake and eddy current brakes are effective at low speed, but the spoilers are much more effective at high speed [8]. Therefore, during normal braking condition, it should be blended several brakes depending on conditions. See Fig.3.

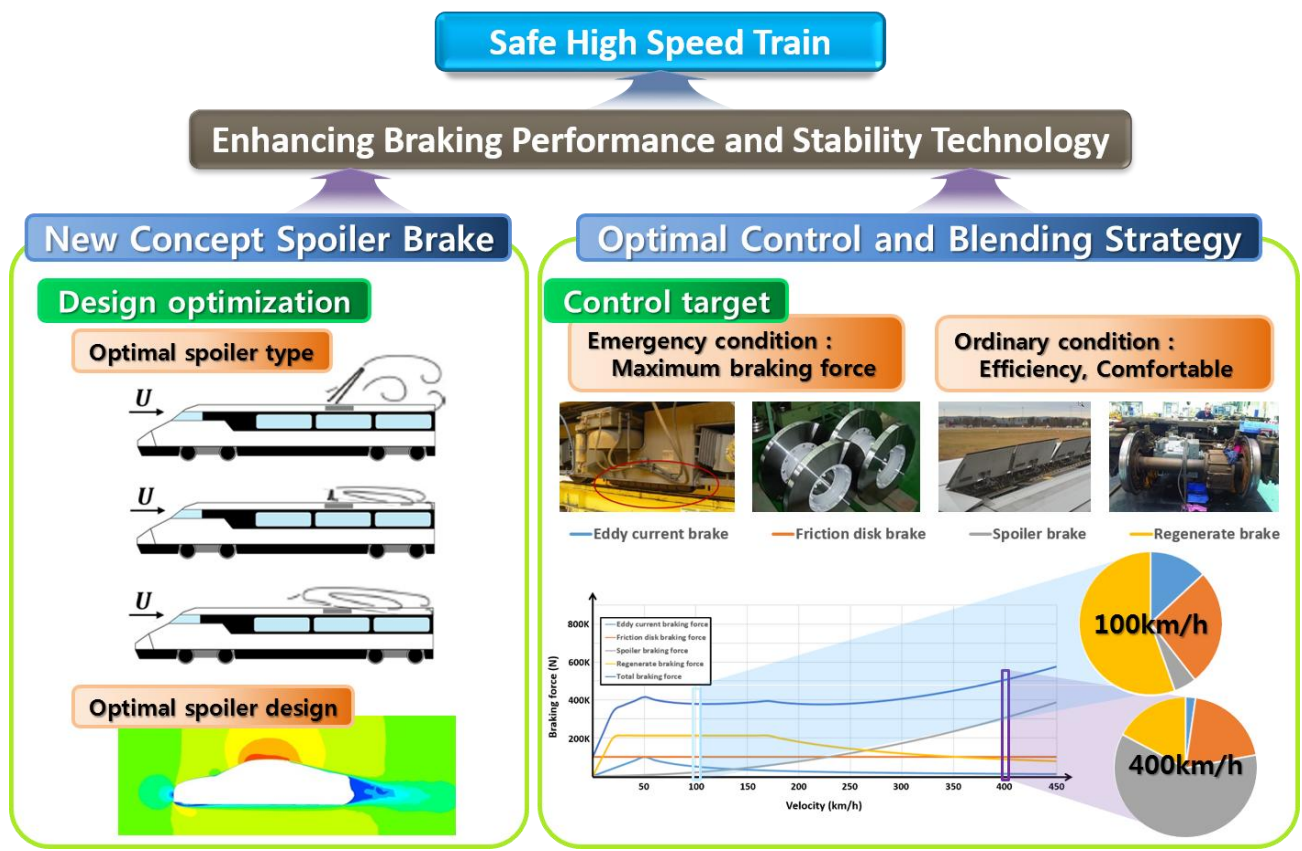

**Figure 3.** Solution approach for enhancing braking performance and stability.

Spoiler brake can introduce unpleasant noise and vibration for passengers, and thus should be carefully controlled and designed. Note that the spoiler design and control have substantial influence on both braking performance and noise and vibration. In addition, the characteristics of each brake differ significantly with the vehicle speed, and thus the brake characteristics of each brake system need to be identified first in order to design a blending strategy [9].

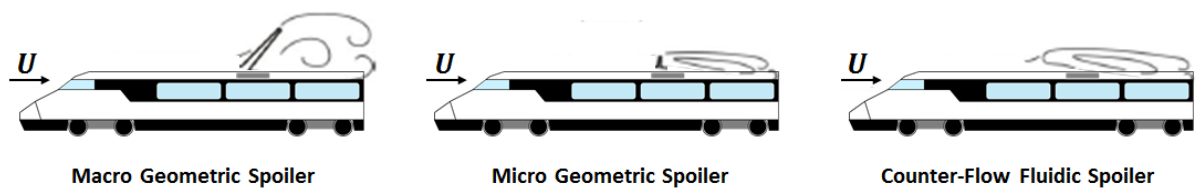

**Figure 4.** Various types and shapes of spoiler.

Braking performance of the spoiler heavily depends on the shape and angle of the spoiler. The aim of this work is to find the best type, shape, and angle of the spoiler for multiple objectives; braking performance, noise/vibrations, and down/lateral forces. Note that these characteristics can change with vehicle speeds. See Fig.4.

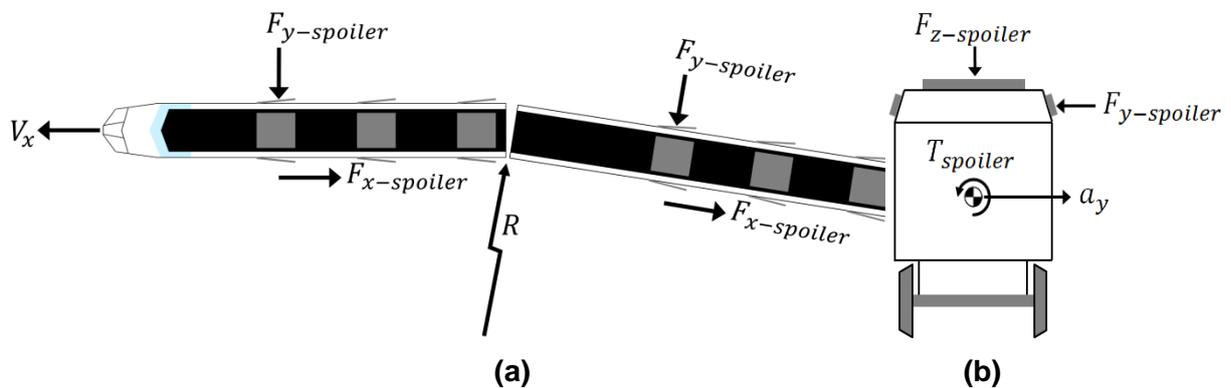

**Figure 5.** (a) Lateral and drag spoiler forces acting on trains during cornering. (b) Down and lateral spoiler forces acting on trains during cornering.

When the emergency braking is used on the curved rail, the train can become unstable due to the braking and lateral forces. Thus, a controller that can prevent the loss of stability with the best braking performance must be designed. In particular, the side spoilers should be properly controlled to generate additional moment force since the behavior of the train depends on its speed and the turning radius [10]. See Fig.5.

By introducing the spoiler brake, multiple braking systems such as disk brake, regenerative brake, eddy current brake, and spoiler brake are available for various braking conditions. Since each brake has different performance and operational characteristics, which varies with speed, their blending strategy must be carefully designed when blending is feasible. For example, in low speed range, the eddy current brake system and regenerate brake have good performance. However in high speed range, the spoiler brake has better performance than the eddy current brake and regenerate brake.

In extremely urgent conditions, all types of brakes should be fully applied. However, in normal braking conditions, we can select and blend the brake depending on brake distance, response time, reliability, efficiency, and passenger comfort.

III. Conclusion

Safety control for various types of trains starts from the interpretation of IoT sensor data. Thus, the IoT sensor data must provide critical information. Therefore, the IoT sensor network should be able to adapt to various situations. Adaption to various situations requires effective operation of IoT sensor network regardless of types of situations. It is also beneficial to system operators of railway industry, because efficient safety control is directly translated into reduction of operational cost. With the additional emergency braking system, the braking distance of the high speed train can be dramatically reduced. The collision due to the braking distance would be prevented. Therefore, the safety of the train operation is guaranteed.

References

[1] A Noureldin, et al., "Performance Enhancement of MEMS-Based INS/GPS Integration for Low-Cost Navigation Applications", IEEE Trans. V.T., vol. 58, no. 3, pp.1077-1096, Mar. 2009.

[2] M Malvezzi, et al., "A localization algorithm for railway vehicles based on sensor fusion between tachometers and inertial measurement units", Proc. of Inst. Mecha. Engine. Part F: Journal of Rail and Rapid Transit, vol.228, no.4, pp.431-448, Apr. 2013.

[3] IEEE 802.15.7(Standard).

[4] Kim H.S, et al., "An Indoor Visible Light Communication Positioning System Using a RF Carrier Allocation Technique", Journal of L.T., vol. 31, no. 1, pp.134-144, Dec. 2012.

[5] Dhahbi S., Abbas-Turki A., "Study of the high-speed trains positioning system: European signaling system ERTMS/ETCS", IEEE Logistics 4[th] Inter. Conf., pp.468-473, May 2011.

[6] I. Park, D. Kim, and D. Har, "MAC Achieving Low Latency and Energy Efficiency in Hierarchical M2M Networks with Clustered Nodes", IEEE Sensor Journal, online published.

[7] E. Sumeu et al., "Modeling and Control of an Eddy Current Brake", Control Eng. Practice, vol. 4, no. 1, pp.19-26, Jan. 1996.

[8] Kapjin Lee et al., "Optimal robust control of a contactless brake system using an eddy current", Mechatronics vol.9, no.6, pp.615-631, Sep. 1999.

[9] P. J. Wang et al., "Analysis of Eddy-current brakes for High Speed Railway", IEEE Transactions on Magnetics, vol. 34, no. 4, pp.1237-1239, Jul. 1998.

[10] G. Naveen Kumar et al., "Dynamic Response of NACA 0018 for Car Spoiler using CFRP Materal", International Journal of Current Engineering and Technology, vol. 4, no. 3, pp.1231-1235, Jun. 2014.